\documentclass[preprint,aps,prl,showpacs,amsmath,amssymb]{revtex4}

\usepackage{graphicx}
\usepackage{dcolumn}
\usepackage{bm}

\begin{document}

\title{New Bardeen-Cooper-Schrieffer-type theory
at finite temperature with particle-number conservation}

\author{H. Nakada}
\email{nakada@faculty.chiba-u.jp}
\affiliation{Department of Physics, Faculty of Science, 
Chiba University, Inage, Chiba 263-8522, Japan}
\author{K. Tanabe}
\email{tanabe@phy.saitama-u.ac.jp}
\affiliation{Department of Physics, Faculty of Science, 
Saitama University, Sakura, Saitama 338-8570, Japan}

\date{\today}

\begin{abstract}
 We formulate a new Bardeen-Cooper-Schrieffer (BCS)-type theory
 at finite temperature,
 by deriving a set of variational equations of the free energy
 after the particle-number projection.
 With its broad applicability,
 this theory can be a useful tool
 for investigating the pairing phase transition
 in finite systems with the particle-number conservation.
 This theory provides effects of the symmetry-restoring fluctuation (SRF)
 for the pairing phenomena in finite fermionic systems,
 distinctively from those of additional quantum fluctuations.
 It is shown by numerical calculations
 that the phase transition is compatible
 with the conservation in this theory,
 and that the SRF shifts up
 the critical temperature ($T^\mathrm{cr}$).
 This shift of $T^\mathrm{cr}$ occurs
 due to reduction of degrees-of-freedom in canonical ensembles,
 and decreases only slowly as the particle-number increases
 (or as the level spacing narrows),
 in contrast to the conventional BCS theory.
\end{abstract}

\pacs{05.70.Fh, 21.60.-n, 74.20.Fg, 05.30.Fk}

\keywords{Finite-temperature BCS theory, Particle-number conservation,
Pairing phase transition}
\maketitle

Pairing phenomena have been observed in various fermionic systems.
Condensate of fermion pairs is realized at low temperature ($T$),
and is usually described
by the Bardeen-Cooper-Schrieffer (BCS) theory~\cite{BCS}
or the Hartree-Fock-Bogolyubov (HFB) theory.
As $T$ increases, the pairing phase transition occurs;
the condensate is dissolved
at a critical temperature $T^\mathrm{cr}$.
In the pairing theories such as BCS or HFB,
the particle-number ($n$) conservation is violated
in the low $T$ (\textit{i.e.} superfluid or superconducting) phase,
as a result of the spontaneous symmetry breaking.
While this picture is reasonable if $n$ is practically infinite,
certain interest has been attracted by finite systems
such as atomic nuclei~\cite{BM2},
in which the condensate of fermion pairs forms at low $T$
but the $n$ conservation is actually preserved.
We have examples also in mesoscopic systems;
\textit{e.g.} granular films~\cite{ZG69},
superconducting islands~\cite{THTT92} and
ultrasmall metallic grains~\cite{RBT,vDR01}.
It has been argued that there are no sharp transitions
in finite systems, whose signatures are washed out
by quantum fluctuations~\cite{MSD72}.
Efforts have been made to find fingerprints
of the superfluid-to-normal phase transition in atomic nuclei.
Recent experiments have revealed an $S$-shape in the graph
of the heat capacity $C$ as a function of $T$,
which was extracted from high-precision
level-density measurements~\cite{Sch01}.
While this $S$-shape was suggested to be the fingerprint,
we have shown that the $n$ conservation produces
a similar $S$-shape even without the transition~\cite{ENT05}.
Under this situation the following fundamental questions are raised
for phase transitions and their relation to finiteness:
(i) for increasing $T$ what roles the conservation law plays
in finite systems, and (ii) how a `phase transition' develops
as $n$ increases.
To answer these questions, it is desired to apply
the $n$ projection in the pairing theory,
particularly in the variation-after-projection (VAP) scheme.
It was shown that exact results in canonical ensembles (CE)
are well approximated by the $n$-projected BCS approach
in the VAP scheme,
for the degenerate model with a constant pairing~\cite{EE93}.
In this Communication, we formulate a new BCS-type theory
at finite temperature with the $n$ projection,
which is well founded on the variational principle
and has much wider applicability.

The BCS theory is closely linked to the Bogolyubov transformation
$c_k^\dagger = u_k \alpha_k^\dagger - v_k \alpha_{\bar{k}}$.
Here $c_k^\dagger$ stands for the creation operator
of the original fermion on the single-particle (s.p.) state $k$,
$\alpha_k^\dagger$ ($\alpha_k$) the creation (annihilation) operator
of quasiparticle (q.p.) on $k$,
and $\bar{k}$ represents the time-reversal to $k$.
The unitarity derives $u_k=\sqrt{1-|v_k|^2}$,
and $v_k$ is usually taken to be real.
At finite $T$, the BCS theory is obtained
for grand-canonical ensembles (GCE),
by assuming the trial statistical operator~\cite{Goo81}
\begin{equation}
w_\mathrm{G}=\frac{e^{-H_0/T}} 
{\mathrm{Tr}(e^{-H_0/T})}\,;\quad 
H_0=\sum_k \varepsilon_k\alpha_k^\dagger \alpha_k\,,
\label{eq:BGop0}
\end{equation}
where we set the Boltzmann constant $k_\mathrm{B}=1$
and the trace is taken over the GCE.
This is an approximation on the excitation spectra of the system
by those of the non-interacting q.p.'s.
The parameters $v_k$ and $\varepsilon_k$,
or equivalently $f_k=1/(e^{\varepsilon_k/T}+1)$,
are determined by variation of the free energy.
This theory is specifically called GCE-BCS theory in this paper.

To include a certain part of $n$ conservation effects,
the number-parity ($\pi_n$) projection~\cite{TSM81} has conveniently
been applied.
However, formulation of the full $n$ projection,
by which we can work in CE,
has been limited for the variation-before-projection (VBP)
scheme~\cite{RR94,TN05},
or for the degenerate model with a constant pairing~\cite{EE93},
except at zero $T$.
Whereas application of the VAP scheme
is desired to investigate effects of a conservation law
as stated earlier,
it has been difficult because of a problem
generic to VAP treatment of spontaneously broken symmetries.
We exemplify the problem for the $n$ projection case.
The $n$ projection operator is written as
\begin{equation}
P_n=\frac{1}{2\pi}\int_0^{2\pi}e^{-i\varphi(N-n)}d\varphi\,,
\label{eq:N-proj}
\end{equation}
where $N=\sum_k N_k$, $N_k = c_k^\dagger c_k$.
With the approximate trial statistical operator
\begin{equation}
w_\mathrm{C}=\frac{P_n e^{-H_0/T}P_n}
 {\mathrm{Tr}(e^{-H_0/T}P_n)}\,,
\label{eq:BGopP}
\end{equation}
the free energy in the BCS-like picture can be defined by
\begin{equation}
F_\mathrm{C}=E_\mathrm{C}-TS_\mathrm{C}\,;\quad
E_\mathrm{C}=\mathrm{Tr}(w_\mathrm{C}H)\,,\quad
S_\mathrm{C}=-\mathrm{Tr}(w_\mathrm{C}\ln w_\mathrm{C})\,.
\label{eq:F_P}
\end{equation}
However, $S_\mathrm{C}$ is not tractable in general cases,
because $[P_n,H_0]\ne 0$ and $\ln w_\mathrm{C}$ leads to
an infinite series of $P_n e^{-H_0/T}P_n$.

To derive a VAP equation in CE,
it is practical to introduce an additional approximation
on the entropy~\cite{TN05},
\begin{equation}
\tilde{F}_\mathrm{C}=E_\mathrm{C}-T\tilde{S}_\mathrm{C}\,;\quad
\tilde{S}_\mathrm{C}=
\frac{1}{T}\frac{\mathrm{Tr}(e^{-H_0/T}H_0 P_n)}
{\mathrm{Tr}(e^{-H_0/T}P_n)}
+\ln\mathrm{Tr}(e^{-H_0/T}P_n)\,.
\label{eq:F_P'}
\end{equation}
The Peierls inequality~\cite{Pei38} $\tilde{F}_\mathrm{C}\geq
F_\mathrm{C}\geq F^\mathrm{exact}$ holds,
which justifies variation of $\tilde{F}_\mathrm{C}$.
Although $\tilde{S}_\mathrm{C}$ can be negative at $T\approx 0$,
violating the third law of thermodynamics,
the variation may suppress influence of this problem.

We here use the following expression,
\begin{equation}
\big[X\big]_\varphi=\frac{\int d\Phi\,X}{\int d\Phi}\,;\quad
 d\Phi=e^{i\varphi(n-\Omega)}\big(\prod_{k>0}\zeta^\varphi_k\big)
 d\varphi\,,
\end{equation}
with $\zeta^\varphi_k=f_k(1-f_{\bar{k}})+(1-f_k)f_{\bar{k}}
 +f_k f_{\bar{k}}\xi_k^{\varphi\ast}
 +(1-f_k)(1-f_{\bar{k}})\xi_k^\varphi$,
$\xi_k^\varphi=u_k^2 e^{i\varphi}+v_k^2 e^{-i\varphi}$,
and $\Omega$ denotes half the number of the s.p. states.
Although $\Omega$ can be infinity in principle,
we should take it to be finite in practical applications,
by introducing a proper cut-off.
The expression $k>0$ for $\prod$ indicates
that the product is taken once for the $(k\bar{k})$ pairs.
We then have, for a given $n$ and $T$,
\begin{eqnarray}
T\delta\tilde{S}_\mathrm{C} &=& \sum_k \varepsilon_k\delta f_k^\mathrm{C}
 + T\sum_{k>0} \big[\frac{\partial\ln\zeta_k^\varphi}
 {\partial v_k}\big]_\varphi\delta v_k\,,\\
\delta E_\mathrm{C} &=& \sum_{k>0}\big[
 h_k^\varphi\delta\rho_k^\varphi
 + h_{\bar{k}}^\varphi\delta\rho_{\bar{k}}^\varphi
 - \Delta_k^\varphi\delta\bar{\kappa}_k^\varphi
 - \bar{\Delta}_k^\varphi\delta\kappa_k^\varphi\big]_\varphi
 + \big[\big(\sum_{k>0}\delta\ln\zeta_k^\varphi\big)\,
 (E^\varphi-E_\mathrm{C})]_\varphi\,,
\end{eqnarray}
where $f_k^\mathrm{C}=\mathrm{Tr}(w_\mathrm{C}\alpha_k^\dagger\alpha_k)$,
$E^\varphi=\mathrm{Tr}(w^\varphi H)$
with $w^\varphi=e^{-i\varphi N}e^{-H_0/T}/
\mathrm{Tr}(e^{-i\varphi N}e^{-H_0/T})$,
$\rho_k^\varphi=\mathrm{Tr}(w^\varphi N_k)$,
$\kappa_k^\varphi=\mathrm{Tr}(w^\varphi c_{\bar{k}} c_k)$,
$\bar{\kappa}_k^\varphi=\mathrm{Tr}(w^\varphi
c_k^\dagger c_{\bar{k}}^\dagger)$, and
\begin{equation}
h_k^\varphi = \frac{\delta E^\varphi}{\delta\rho_k^\varphi}\,,\quad
-\Delta_k^\varphi = \frac{\delta E^\varphi}
{\delta\bar{\kappa}_k^\varphi}\,,\quad
-\bar{\Delta}_k^\varphi = \frac{\delta E^\varphi}
{\delta\kappa_k^\varphi}\,.
\end{equation}
Owing to the extended Wick's theorem~\cite{BB69},
$E^\varphi$ can be expressed in terms of
$\rho_k^\varphi$, $\kappa_k^\varphi$ and $\bar{\kappa}_k^\varphi$.
The variation of $\tilde{F}_\mathrm{C}$ with respect to $f_k$
leads to the following coupled equations of $\varepsilon_k$,
\begin{eqnarray}
\sum_{k'} \varepsilon_{k'}\frac{\partial f_{k'}^\mathrm{C}}{\partial f_k}
 &=& \left[\big\{h_k^\varphi-(h_k^\varphi+h_{\bar{k}}^\varphi)
 \big(f_{\bar{k}}-(f_{\bar{k}}-v_k^2)e^{-i\varphi}\big)
 +u_k v_k(\Delta_k^\varphi e^{-i\varphi}
 +\bar{\Delta}_k^\varphi e^{i\varphi})\big\}/\zeta_k^\varphi\right]_\varphi
 \nonumber\\
 && - \big[(D_k^\varphi-E^\varphi+E_\mathrm{C})
  \frac{\partial\ln\zeta_k^\varphi}{\partial f_k}\big]_\varphi\,,
 \label{eq:qp-eq}
\end{eqnarray}
where $D_k^\varphi =  h_k^\varphi\rho_k^\varphi
 + h_{\bar{k}}^\varphi\rho_{\bar{k}}^\varphi
 - \Delta_k^\varphi\bar{\kappa}_k^\varphi
 - \bar{\Delta}_k^\varphi\kappa_k^\varphi$.
Since $f_k^\mathrm{C} = f_k + f_k(1-f_k)
 \big[\frac{\partial\ln\zeta_k^\varphi}{\partial f_k}\big]_\varphi$,
for the lhs of Eq.~(\ref{eq:qp-eq}) we have
\begin{eqnarray}
\frac{\partial f_{k'}^\mathrm{C}}{\partial f_k} &=&
 \delta_{kk'}\left\{1+(1-2f_k)
 \big[\frac{\partial\ln\zeta_k^\varphi}{\partial f_k}\big]_\varphi\right\}
 -2\delta_{\bar{k}k'}f_{\bar{k}}(1-f_{\bar{k}})
 \big[\frac{1-\cos\varphi}{\zeta_k^\varphi}\big]_\varphi\nonumber\\
 && +f_{k'}(1-f_{k'})\left\{(1-\delta_{kk'}-\delta_{\bar{k}k'})
  \big[\frac{\partial\ln\zeta_k^\varphi}{\partial f_k}
  \frac{\partial\ln\zeta_{k'}^\varphi}{\partial f_{k'}}\big]_\varphi
 - \big[\frac{\partial\ln\zeta_k^\varphi}{\partial f_k}\big]_\varphi
 \big[\frac{\partial\ln\zeta_{k'}^\varphi}{\partial f_{k'}}\big]_\varphi
 \right\}\,.
\end{eqnarray}
The variation of $\tilde{F}_\mathrm{C}$ with respect to $v_k$ yields
\begin{eqnarray}
&&\hspace*{-1.5cm}
 2u_k v_k\tilde{h}_k - (u_k^2-v_k^2)\tilde{\Delta}_k = 0\,;
 \label{eq:CE-BCS}\\
\tilde{\Delta}_k &=& \frac{1}{2}\big[(\Delta_k^\varphi e^{-i\varphi}
 +\bar{\Delta}_k^\varphi e^{i\varphi})\}/\zeta_k^\varphi\big]_\varphi\,,
 \label{eq:CE-gap}\\
\tilde{h}_k &=& \big[\frac{1}{2}(h_k^\varphi+h_{\bar{k}}^\varphi)
 \frac{e^{-i\varphi}}{\zeta_k^\varphi}
 + (D_k^\varphi-E^\varphi+E_\mathrm{C}+T)
 \frac{i\sin\varphi}{\zeta_k^\varphi}\big]_\varphi
 \nonumber\\
 && -\sum_{k'} \varepsilon_{k'}f_{k'}(1-f_{k'})\,\left[\bigg\{
  \frac{\delta_{kk'}+\delta_{\bar{k}k'}}{1-f_k-f_{\bar{k}}}
  -(1-\delta_{kk'}-\delta_{\bar{k}k'})
  \frac{\partial\ln\zeta_{k'}^\varphi}{\partial f_{k'}}
  +\big[\frac{\partial\ln\zeta_{k'}^{\varphi'}}{\partial f_{k'}}
  \big]_{\varphi'} \bigg\}\right.\nonumber\\
 &&\hspace*{10cm} \left.\times\frac{i\sin\varphi}{\zeta_k^\varphi}
  \right]_\varphi\,. \label{eq:CE-spe}
\end{eqnarray}
Equation~(\ref{eq:CE-BCS}) can be solved
in analogy to the usual BCS equation,
\begin{equation}
v_k^2 = \frac{1}{2}\bigg(1-\frac{\tilde{h}_k}
 {\sqrt{\tilde{h}_k^2+\tilde{\Delta}_k^2}}\bigg)\,.
\label{eq:CE-v2}\end{equation}
In this regard $\tilde{\Delta}_k$ plays a similar role
to the gap parameter in the usual BCS theory.

We shall call the above formalism
\textit{canonical-ensemble BCS (CE-BCS) theory}.
The CE-BCS theory recovers the $n$-projected BCS theory
for the ground state in the $T\rightarrow 0$ limit,
and coincides the GCE-BCS theory
when $d\Phi$ is replaced by $\delta(\varphi)\,d\varphi$.
From the GCE-BCS theory viewpoint,
the $\varphi$ variable in Eq.~(\ref{eq:N-proj})
corresponds to the Nambu-Goldstone mode
which the broken symmetry gives rise to.
The $\varphi$ integration takes account of
the quantum fluctuation of the Nambu-Goldstone field.
Since the $n$ conservation is restored by this fluctuation,
we call this fluctuation \textit{symmetry-restoring fluctuation (SRF)}.
Via the $n$ projection the SRF is separated
from additional quantum fluctuations (AQF).
The SRF may change configuration (\textit{e.g.} $v_k$)
that minimizes the free energy.
This effect is taken into account in the CE-BCS framework,
not in the VBP scheme.
It is also noted that the CE-BCS theory is broadly applicable
with no limitation on Hamiltonian,
and is relatively easy to increase $n$
or to enlarge the model space,
compared with other extensions of the BCS theory
that restore the $n$ conservation.

We present a numerical application of the CE-BCS theory.
The following Hamiltonian is adopted for the sake of simplicity,
\begin{equation}
 H = \sum_k (t_k-\mu) N_k - g\,B^\dagger B\,;\quad
 B = \sum_{k>0} c_{\bar{k}} c_k\,.
\label{eq:pr-H}\end{equation}
In any of the examples below,
$g$ is adjusted so as for the gap parameter
of the GCE-BCS approximation (which is denoted by $\Delta_\mathrm{G}$)
to be unity at zero $T$.
In other words, all quantities having the energy dimension
are represented in unit of $\Delta_\mathrm{G}(T=0)$.
The model space is cut off by $|t_k|\leq\Lambda$,
and we set $\Lambda=10$.
The parameter $t_k$ is chosen to be
$t_k=-\Lambda+(k-1)d$ where $d=2\Lambda/(\Omega-1)$
$(k=1, 2, \cdots, \Omega)$,
with the time-reversal symmetry $t_{\bar{k}}=t_k$.
Keeping $n=\Omega$ (\textit{i.e.} half-filled),
the particle number $n$ is varied.
In this paper we restrict ourselves to $n=\mbox{even}$ cases.
The parameter $\mu$ in Eq.~(\ref{eq:pr-H}) merely shifts
the zero-point of energy in CE,
while it corresponds to the particle number condition in GCE.
Indeed, though not trivial in Eqs.~(\ref{eq:qp-eq},\ref{eq:CE-spe}),
the CE-BCS results hardly depend on $\mu$.
For fast and safe convergence in numerical calculations,
minimization of $\tilde{F}_\mathrm{C}$
is implemented by combining Eqs.~(\ref{eq:qp-eq},\ref{eq:CE-v2})
with the steepest decent method.

The approximation on the entropy by $\tilde{S}_\mathrm{C}$
has been tested for the $n=26$ case.
Since it is difficult to compute $S_\mathrm{C}$,
we compare $\tilde{S}_\mathrm{C}$ with the exact canonical entropy
$S^\mathrm{exact}=-\mathrm{Tr}(w\ln w)$ where $w=P_n e^{-H/T}P_n
/\mathrm{Tr}(e^{-H/T}P_n)$, which can be obtained
by the quantum Monte Carlo (QMC) calculation~\cite{SMMC},
using the method described in Ref.~\cite{NA97}.
Although there is a slight discrepancy at low $T$
as mentioned earlier
and a weak kink around $T=T^\mathrm{cr}_\mathrm{C}$
($T^\mathrm{cr}_\mathrm{C}$ will be defined later),
$\tilde{S}_\mathrm{C}$ is found to be in moderate agreement
with $S^\mathrm{exact}$ at any $T$.

\begin{figure}
\includegraphics[scale=0.75]{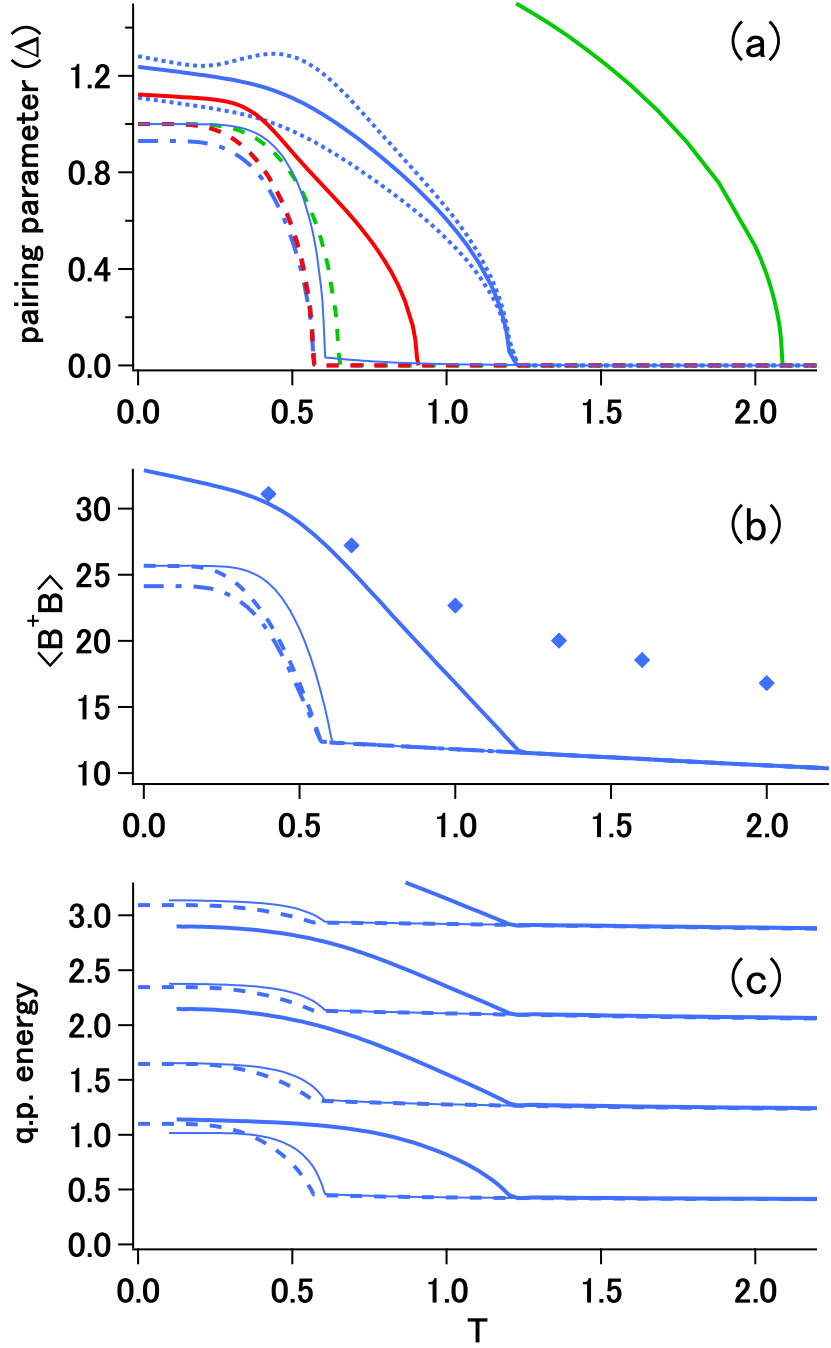}%
\caption{(a) $n$- and $T$-dependence of pairing parameters.
 Colors distinguish $n$; green for $n=10$, blue for $n=26$
 and red for $n=56$.
 For each color the thick solid line represents
 $\Delta^\mathrm{av}_\mathrm{C}$,
 while the dashed line $\Delta_\mathrm{G}$.
 For the $n=26$ case, $\tilde{\Delta}_\mathrm{min}$ and
 $\tilde{\Delta}_\mathrm{max}$ (dotted lines),
 $\Delta^\mathrm{av}_{\mathrm{C}'}$ (dot-dashed line)
 and $\Delta^\mathrm{av}_\pi$ (thin solid line) are also presented.
 (b) $\langle B^\dagger B\rangle$ for $n=26$.
 Diamonds are the exact quantum Monte Carlo results.
 We use the same conventions for lines as in (a).
 (c) $\varepsilon_k$ ($k-n/2=1,2,3,4$) for $n=26$.
 Conventions are the same as in (a).
 \label{fig:Delta}}
\end{figure}

The thermal expectation value of an operator $O$
in the CE-BCS (GCE-BCS) is denoted by $\langle O\rangle_\mathrm{C}$
($\langle O\rangle_\mathrm{G}$).
Replacing $P_n$ by the $\pi_n$ projection operator~\cite{TN05},
we can calculate $\pi_n$-projected counterparts
to the $n$-projected quantities.
The $\pi_n$-projected expectation value of $O$
will be considered for comparison,
which is expressed as $\langle O\rangle_\pi$,
as well as the expectation value in the VBP
$\langle O\rangle_{\mathrm{C}'}(=\langle O P_n\rangle_\mathrm{G})$.
The gap parameter $\Delta_\mathrm{G}(=g\langle B\rangle_\mathrm{G})$
is regarded as an order parameter for the pairing transition
in the GCE-BCS theory.
However, obviously $\langle B\rangle_\mathrm{C}=
\langle B\rangle_{\mathrm{C}'}=0$ at any $T$.
Instead we consider two alternative definitions;
one is $\tilde{\Delta}_k$ in Eq.~(\ref{eq:CE-gap}),
and the other is $\Delta^\mathrm{av}\equiv g \sqrt{
\langle B^\dagger B\rangle - \sum_{k>0} \langle N_k\rangle\,
\langle N_{\bar{k}}\rangle}$~\cite{vDR01}.
For the latter,
we define $\Delta^\mathrm{av}_\mathrm{C}$,
$\Delta^\mathrm{av}_{\mathrm{C}'}$ and so forth,
in accordance with the expression for $\langle~\rangle$.
These pairing parameters have lost
direct connection to the energy gap.
In Fig.~\ref{fig:Delta}(a),
$\Delta_\mathrm{G}(T)$ and $\Delta^\mathrm{av}_\mathrm{C}(T)$
are depicted for various $n$ values.
For comparison, $\Delta^\mathrm{av}_{\mathrm{C}'}(T)$
and $\Delta^\mathrm{av}_\pi(T)$ are presented for the $n=26$ case.
Because $\tilde{\Delta}_k$ depends on $k$,
$\tilde{\Delta}_\mathrm{min}(T)\equiv
\min_k\big(\tilde{\Delta}_k(T)\big)$
and $\tilde{\Delta}_\mathrm{max}(T)\equiv
\max_k\big(\tilde{\Delta}_k(T)\big)$ are shown for $n=26$.
It is confirmed that $\tilde{\Delta}_k$'s and
$\Delta^\mathrm{av}_\mathrm{C}$ do not differ much from one another.

The gap parameter $\Delta_\mathrm{G}$ vanishes
at the critical temperature $T^\mathrm{cr}_\mathrm{G}(\approx 0.6)$
in the GCE-BCS theory.
As $n$ increases $\Delta_\mathrm{G}$ converges rapidly;
$\Delta_\mathrm{G}$'s for $n=26$ and $56$ cannot be distinguished
in Fig.~\ref{fig:Delta}(a).
It is remarked that, in the CE-BCS results,
transition-like behavior remains.
$\Delta^\mathrm{av}_\mathrm{C}$ vanishes at a certain temperature
$T^\mathrm{cr}_\mathrm{C}$,
which is appreciably higher than $T^\mathrm{cr}_\mathrm{G}$.
$T^\mathrm{cr}_\mathrm{C}$ approaches $T^\mathrm{cr}_\mathrm{G}$
as $n$ grows, but only gradually.
Although shift of the critical temperature is already found
in $\Delta^\mathrm{av}_\pi$ (giving $T^\mathrm{cr}_\pi$),
it is far less significant than in $\Delta^\mathrm{av}_\mathrm{C}$.
While the SRF enhances the pairing parameter in the CE-BCS results,
the VBP scheme gives smaller $\Delta^\mathrm{av}_{\mathrm{C}'}$
than $\Delta_\mathrm{G}$.
Effects of the SRF are not carried correctly in the VBP,
since the configuration change is discarded.
In Fig.~\ref{fig:Delta}(b),
we compare the $\langle B^\dagger B\rangle$ values
in the BCS-type approximations with the exact ones for $n=26$,
which are evaluated by the QMC calculation.
Taking account of the SRF effects,
$\langle B^\dagger B\rangle_\mathrm{C}$'s are close to the exact values
at $T<T^\mathrm{cr}_\mathrm{C}$,
unlike $\langle B^\dagger B\rangle_\mathrm{G}$,
$\langle B^\dagger B\rangle_{\mathrm{C}'}$
and $\langle B^\dagger B\rangle_\pi$.
Thus the CE-BCS theory gives significant improvement
over the GCE-BCS and the $\pi_n$-projected theories,
while keeping simplicity of the BCS picture to a considerable extent.
The deviation at $T\gtrsim T^\mathrm{cr}_\mathrm{C}$
may be attributed mainly to influence of the AQF.

The above behavior of the pairing parameters
is reflected in the q.p. energies.
Although $\varepsilon_k$'s are parameters independent of $v_k$'s,
they also correlate well to the pairing parameters,
having a certain energy gap at $T<T^\mathrm{cr}_\mathrm{C}$,
as shown in Fig.~\ref{fig:Delta}(c).
Since $\varepsilon_k$ is expected to give an approximate energy
of the adjacent odd-$n$ system,
this suggests even-odd difference in energy
at $T<T^\mathrm{cr}_\mathrm{C}$.

\begin{figure}
\includegraphics[scale=0.75]{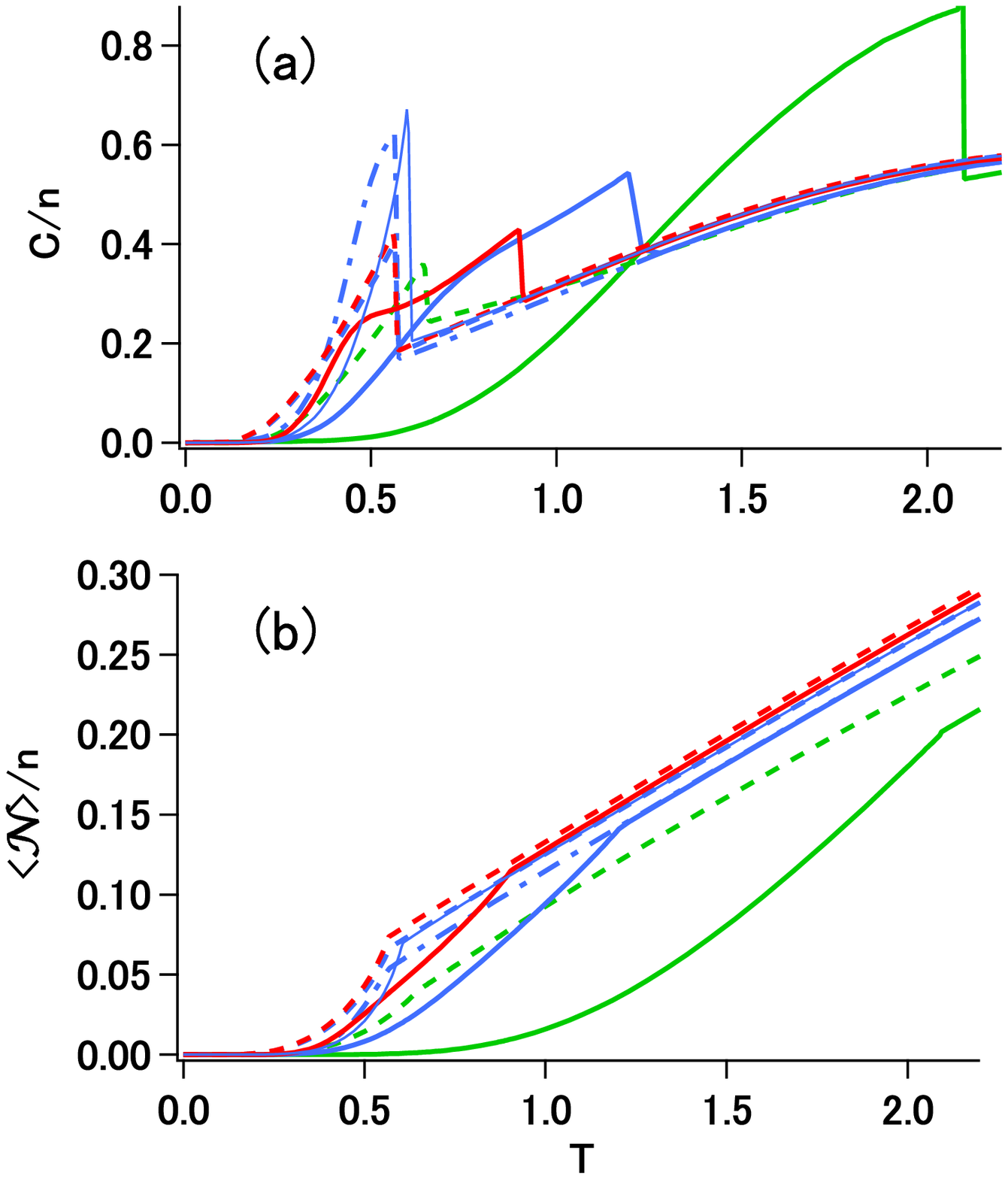}%
\caption{$n$- and $T$-dependence of (a) specific heat $C/n$,
 and (b) $\langle\mathcal{N}\rangle/n$.
 See Fig.~\ref{fig:Delta} for conventions.
 \label{fig:C}}
\end{figure}

We next view the heat capacity $C(T)=dE/dT$ ($C_\mathrm{G}$ for GCE-BCS,
$C_\mathrm{C}$ for CE-BCS, and so forth),
whose singular structure, if any,
is linked to a phase transition in general.
Figure~\ref{fig:C}(a) shows a specific heat $C(T)/n$,
where $C$ is computed by numerical differentiation
of $E=\langle H\rangle$.
As $C_\mathrm{G}$ at $T^\mathrm{cr}_\mathrm{G}$,
$C_\mathrm{C}$ has discontinuity at $T^\mathrm{cr}_\mathrm{C}$.
It is thus fair to say that the pairing transition remains
in the CE-BCS approximation, but at $T^\mathrm{cr}_\mathrm{C}$
that is substantially higher than $T^\mathrm{cr}_\mathrm{G}$.
The SRF shifts up $T^\mathrm{cr}$, not erasing the transition,
in the CE-BCS theory.
The $S$-shape behavior of $C(T)$ at low $T$
may be compared to those observed~\cite{Sch01}.
Comparison with the exact QMC result
draws consistent consequence with the $\langle B^\dagger B\rangle$ case
in Fig.~\ref{fig:Delta}(b).

In Fig.~\ref{fig:C}(b), we present expectation values
of the q.p. number $\mathcal{N}=\sum_k \alpha_k^\dagger \alpha_k$,
as a function of $T$.
It is found that the shift from $T^\mathrm{cr}_\mathrm{G}$
to $T^\mathrm{cr}_\mathrm{C}$ is closely connected
to the difference between $\langle\mathcal{N}\rangle_\mathrm{G}$
and $\langle\mathcal{N}\rangle_\mathrm{C}$.
The $n$ projection reduces the number of states.
At low $T$, lack of one q.p. states gives rise to
reduction of $\langle\mathcal{N}\rangle_\pi$
and $\langle\mathcal{N}\rangle_\mathrm{C}$,
in comparison with $\langle\mathcal{N}\rangle_\mathrm{G}$~\cite{ENT05}.
At higher $T$, the $n$ projection eliminates
some of the higher q.p. degrees-of-freedom.
Therefore, $\langle\mathcal{N}\rangle_\mathrm{C}$
becomes smaller than $\langle\mathcal{N}\rangle_\pi$.
Moreover, as $T$ becomes higher,
the entropy increases more slowly in CE
than in GCE and in the $\pi_n$-projected space.
This further delays rise of $\langle\mathcal{N}\rangle_\mathrm{C}$
through the configuration change.
This effect is the stronger for the fewer $n$,
and leads to the shift of $T^\mathrm{cr}$.

\begin{figure}
\includegraphics[scale=0.85]{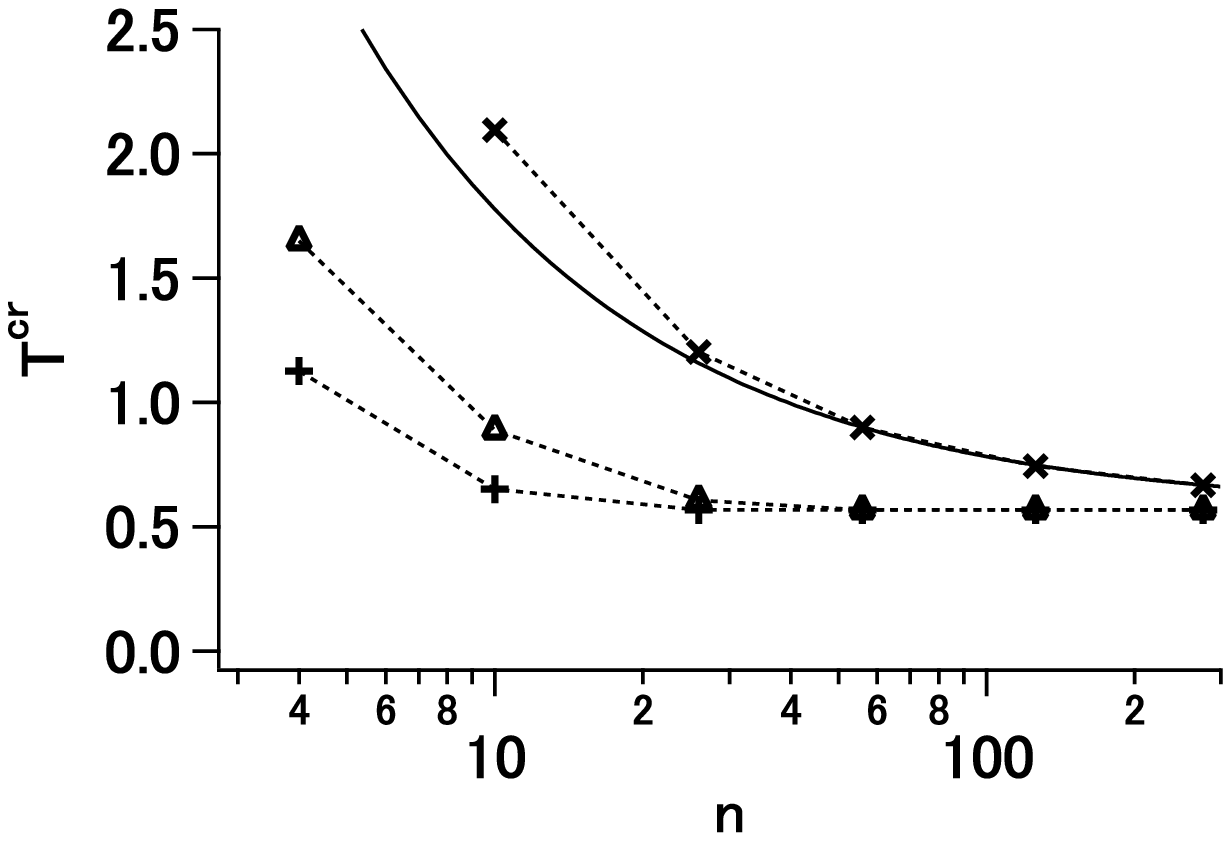}%
\caption{$n$-dependence of $T^\mathrm{cr}$;
 $T^\mathrm{cr}_\mathrm{G}$ (pluses), $T^\mathrm{cr}_\pi$ (triangles)
 and $T^\mathrm{cr}_\mathrm{C}$ (crosses).
 The solid line is a fit to $T^\mathrm{cr}_\mathrm{C}$;
 $T^\mathrm{cr}=T^\mathrm{cr}_\infty + 6.8\,n^{-0.75}$.
 \label{fig:Tc}}
\end{figure}

Figure~\ref{fig:Tc} depicts the $n$-dependence of $T^\mathrm{cr}$.
$T^\mathrm{cr}_\mathrm{G}$ rapidly goes to the bulk
(\textit{i.e.} $n\rightarrow\infty$) limit $T^\mathrm{cr}_\infty$.
On the contrary $T^\mathrm{cr}_\mathrm{C}$
approaches $T^\mathrm{cr}_\infty$ slowly.
We find, by fitting, $T^\mathrm{cr}_\mathrm{C}-T^\mathrm{cr}_\infty
\propto n^{-0.75}\propto d^{0.75}$, except for quite small $n$.
For a fixed s.p. level spacing $d$,
$T^\mathrm{cr}_\mathrm{C}$ is insensitive to the cut-off $\Lambda$,
although it slightly goes up as $\Lambda$ increases.

The `phase transition' picture originates from the approximation
of $H$ by $H_0$ in the trial statistical operator.
While the transition will be washed out due to the AQF,
the present study yields the approximate picture
in which the phase transition is compatible with the conservation law.
The finite system approaches the bulk limit
as both the SRF and the AQF are reduced,
and the reduction of the SRF gives rise to
the decrease of $T^\mathrm{cr}_\mathrm{C}$.
It should be commented, however,
that in the degenerate model~\cite{EE93}
the exact treatment of $S_\mathrm{C}$ was shown to wash out
the signatures of the transition,
while the CE-BCS theory keeps the transition behavior
due to the approximation of $S_\mathrm{C}$ by $\tilde{S}_\mathrm{C}$.
Even slight difference in the entropy could influence
occurrence of the transition.
Whereas the present study gives a new insight,
further investigation is necessary for full understanding
of the superfluid-to-normal transition under the $n$ conservation.

In summary,
we have formulated a BCS-type theory in canonical ensembles
(CE-BCS theory) at finite temperature,
by introducing an approximation on the entropy
with retaining the variational principle.
The CE-BCS theory keeps some of the structure
in the conventional grand-canonical ensemble BCS (GCE-BCS) theory.
The equations in the theory are well connected to the GCE-BCS theory,
so that, under the particle-number conservation,
the GCE-BCS theory could be regarded
as an approximation of the CE-BCS theory.
Having broad applicability to a moderately good accuracy,
the CE-BCS theory provides us with a useful tool to investigate
effects of the symmetry-restoring fluctuation (SRF),
separately from those of the additional quantum fluctuations,
although a part of the SRF might be missed
due to the approximation on the entropy.
Numerical application of this theory gives a new picture
for the pairing transition in finite systems,
in which the pairing transition is reconciled
with the particle-number conservation.
In the CE-BCS framework the SRF has been found to shift up
the critical temperature $T^\mathrm{cr}$.
The shift of $T^\mathrm{cr}$ occurs due to the reduction
of excitation degrees-of-freedom in canonical ensembles.
$T^\mathrm{cr}$ in the CE-BCS theory
approaches its bulk limit much more slowly
than that in the GCE-BCS theory,
for narrowing single-particle level spacing.
Several more aspects of the CE-BCS theory
that give us significant insights into the pairing phenomena
will be discussed in a future publication.

\begin{acknowledgments}
The present work is financially supported in part
as Grant-in-Aid for Scientific Research (B), No.~15340070,
by the MEXT, Japan.
Some of the numerical calculations are performed on HITAC SR11000
at Institute of Media and Information Technology, Chiba University,
and at Information Technology Center, University of Tokyo.
\end{acknowledgments}

\end{document}